\documentclass[prc,preprint,showpacs,preprintnumbers,amsmath,amssymb]{revtex4}

\usepackage[mathscr]{eucal}
\usepackage{graphicx}
\def\slr#1{\setbox0=\hbox{$#1$}           
   \dimen0=\wd0                                 
   \setbox1=\hbox{/} \dimen1=\wd1               
   \ifdim\dimen0>\dimen1                        
      \rlap{\hbox to \dimen0{\hfil/\hfil}}      
      #1                                        
   \else                                        
      \rlap{\hbox to \dimen1{\hfil$#1$\hfil}}   
      /                                         
   \fi}

\def\mytint#1{\!\int\!\!\frac{d^3\!{#1}}{(2\pi)^3}\,}

\def\be{\begin{eqnarray}}
\def\ee{\end{eqnarray}}

\begin{document}

\preprint{BCCNT: 05/12/323}

\title{Calculation of Screening Masses in a Chiral Quark Model}

\author{Xiangdong Li}
\affiliation{%
Department of Computer System Technology\\
New York City College of Technology of the City University of New
York\\
Brooklyn, New York 11201 }%

\author{Hu Li}
\author{C. M. Shakin}
\email[email address:]{casbc@cunyvm.cuny.edu}
\author{Qing Sun}

\affiliation{%
Department of Physics and Center for Nuclear Theory\\
Brooklyn College of the City University of New York\\
Brooklyn, New York 11210
}%

\date{May, 2004}

\begin{abstract}
We consider a simple model for the coordinate-space vacuum
polarization function which is often parametrized in terms of a
screening mass. We discuss the circumstances in which the standard
result for the screening mass , $m_{sc}=\pi T$, is obtained. In
the model considered here, that result is obtained when the
momenta in the relevant vacuum polarization integral are small
with respect to the first Matsubara frequency.
\end{abstract}


\maketitle

 In a number of recent works [1-\,3] we have calculated
various hadronic correlation functions and compared our results to
results obtained in lattice simulations of QCD [4-\,6]. The
lattice results for the correlators, $G(\tau, T)$, may be used to
obtain the corresponding spectral functions, $\sigma(\omega, T)$,
by making use of the relation \be G(\tau, T)=\int_0^\infty d
\omega \sigma(\omega, T) K(\tau, \omega, T)\,,\ee where \be
K(\tau, \omega,
T)=\frac{\cosh[\omega(\tau-1/2T)]}{\sinh(\omega/2T)}\,.\ee The
procedure to obtain $\sigma(\omega, T)$ from the knowledge of
$G(\tau, T)$ makes use of the maximum entropy method (MEM) [7-9],
since $G(\tau, T)$ is only known at a limited number of points.

In our studies of meson spectra at $T=0$ and at $T<T_c$ we have
made use of the Nambu--Jona-Lasinio (NJL) model. The Lagrangian of
the generalized NJL model we have used in our studies is

\begin{flushleft}
\be \mathcal L&=&\overline{q}(i\slr\gamma-m^0)q+\frac{\overline
G_S}{2}\sum_{i=0}^8 [(\overline{q} \lambda^{i} q)^2+(\overline{q}
i \gamma_5 \lambda^{i} q)^2]\\\nonumber &-&\frac{\overline
G_V}{2}\sum_{i=0}^{8}[(\overline{q} \lambda^{i}\gamma_\mu
q)^2+(\overline{q} \lambda^{i}\gamma_5\gamma_\mu q)^2]\\\nonumber
&+&\frac{G_D}{2} \lbrace
\det[\overline{q}(1+\lambda_5)q]+\det[\overline{q}(1-\lambda_5)q]\rbrace
+\mathcal L_{conf}\,. \ee
\end{flushleft}

Here, $m^0$ is a current quark mass matrix, $m^0=diag(m_u^0,
m_d^0, m_s^0)$. The $\lambda_i$ are the Gell-Mann (flavor)
matrices and $\lambda^0=\sqrt{2/3}\mathbf{1}$, with $\mathbf{1}$
being the unit matrix. The fourth term is the 't Hooft interaction
and $\mathcal L_{conf}$ represents the model of confinement used
in our studies of meson properties.

In the study of hadronic current correlators it is important to
use a model which respects chiral symmetry, when $m^0=0$.
Therefore, we make use of the Lagrangian of Eq.\,(3), while
neglecting the 't Hooft interaction and $\mathcal L_{conf}$. In
order to make contact with the results of lattice simulations we
use the model with the number of flavors, $N_f=1$. Therefore, the
$\lambda^i$ matrices in Eq.\,(3) may be replaced by unity. We then
have used \be \mathcal
L&=&\overline{q}(i\slr\gamma-m^0)q+\frac{G_S}{2}[(\overline{q}q)^2+(\overline{q}
i \gamma_5 q)^2]\\\nonumber &-&\frac{G_V}{2}[(\overline{q}
\gamma_\mu q)^2+(\overline{q}\gamma_5\gamma_\mu q)^2] \ee in order
to calculate the hadronic current correlation functions in earlier
work [1-3].

    In order to present our results in the simplest form, we
consider only the scalar interaction proportional to
$(\overline{q}q)^{2}$. We also extend the definition of
$\sigma(\omega,T)$ of Eq.\,(1) to include a dependence upon the
total moment of the quark and antiquark appearing in the
polarization integral. Thus we consider the imaginary part of the
correlator, $\sigma(\omega,\overrightarrow{P})$. Since we place
$\overrightarrow P$ along the \emph{z}-axis this quantity may be
written as $\sigma(\omega, 0, 0, P_z)$ in accord with the notation
of Ref. [10].In this work we will present our result for the
coordinate-dependent correlator $C(z)$ which is proportional to
the correlator defined in Eq.\,(1) of Ref. [10], \be
C(z)=\frac12\int_{-\infty}^\infty\,dP_ze^{iP_zz}\int_0^\infty\,d\omega\frac{\sigma(\omega,
0, 0, P_z)}\omega\,. \ee We may also use the form \be
C(z)=\frac14\int_{-\infty}^\infty\,dP_ze^{iP_zz}\int_0^\infty\,dP^2\,\frac{\sigma(P^2,
0, 0, P_z)}{P^2}\,. \ee

We have made a study of the screening mass in a simple model in
order to understand the origin of exponential behavior for the
correlator. To that end we make use of Ref. [11]. We consider the
Matsubara formalism and note that the quark propagator may be
written, with $\beta=1/T$, \be
S_\beta(\overrightarrow{k},\omega_n)=\frac{\gamma^0(2n+1)\pi/\beta+
\overrightarrow{\gamma}\cdot\overrightarrow{k}-M}{(2n+1)^2\pi^2/\beta^2+\overrightarrow{k}^2+M^2}\,.\ee
For bosons the vacuum polarization function  is given as Eq.
(1.51) of Ref. [11],
\be\Pi(\overrightarrow{p},p^0)=\frac{g^2}{2\beta}\sum_n\frac{d^3k}{(2\pi)^3}\,
\frac1{\dfrac{4n^2\pi^2}{\beta^2}+\overrightarrow{k}^2+M^2}\cdot\frac{1}
{\left(\dfrac{2n\pi}{\beta}+p^0\right)^2+(\overrightarrow{k}+\overrightarrow{p})^2+M^2}\,.\ee

We modify Eq.\,(8) to refer to fermions. In this case the
Matsubara frequencies are \be\omega_n=\frac{(2n+1)\pi}\beta\ee and
we have
\be\Pi(\overrightarrow{p},p^0)=\frac{g^2}{2\beta}\mbox{Tr}\mytint
k\frac{\left[\left(\gamma^0\pi/\beta+\overrightarrow{\gamma}\cdot\overrightarrow{k}\right)
\left(\gamma^0(p^0+\pi/\beta)+\overrightarrow{\gamma}\cdot(\overrightarrow{k}+\overrightarrow{p})\right)\right]}
{\left(\dfrac{\pi^2}{\beta^2}+\overrightarrow{k}^2\right)\left[\left(\dfrac\pi\beta+p^0\right)^2+
(\overrightarrow{k}+\overrightarrow{p})^2\right]}\,,\ee if we keep
only the first term in the sum, where $\omega_0=\pi/\beta$. As a
next step we drop $p^0$, so that we have
\be\Pi(\overrightarrow{p},0)=\frac{g^2}{2\beta}\mbox{Tr}\mytint
k\frac{\left[\left(\gamma^0\pi/\beta+\overrightarrow{\gamma}\cdot\overrightarrow{k}\right)
\left(\gamma^0\pi/\beta+\overrightarrow{\gamma}\cdot(\overrightarrow{k}+\overrightarrow{p})\right)\right]}
{\left(\left(\dfrac{\pi}{\beta}\right)^2+\overrightarrow{k}^2\right)\left[\left(\dfrac\pi\beta\right)^2+
(\overrightarrow{k}+\overrightarrow{p})^2\right]}\,.\ee We then
take $\overrightarrow{p}$ along the $z$ axis and write
$\Pi(p_z)=\Pi(\overrightarrow{p},0)$. We define \be C(z)=\int
dp_z\,e^{ip_zz}\,\Pi(p_z)\,.\ee In our calculation we replace
$g^2/2\beta$ by unity and use a sharp cutoff so that
$|\overrightarrow{k}|<k_{max}$.

\begin{figure}
\includegraphics[bb=0 0 280 235, angle=0, scale=1]{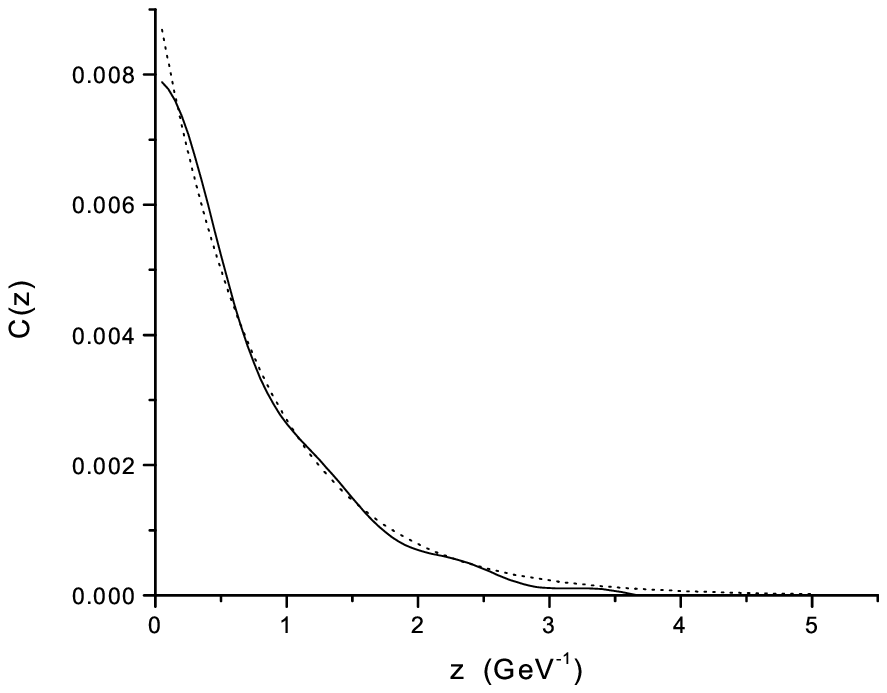}%
\caption{The function $C(z)$ of Eq.\,(6) is shown for a sharp
cutoff of $k_{max}=0.1$ GeV. The dotted line represents an
exponential fit to the curve using $m_{sc}=1.23$ GeV. (We recall
that $\pi T$ is equal to 1.27 GeV.)}
\end{figure}

\begin{figure}
\includegraphics[bb=0 0 280 235, angle=0, scale=1]{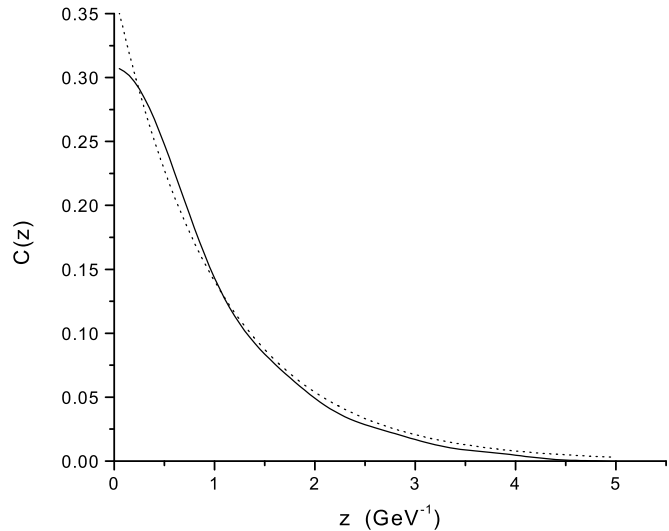}%
\caption{The function $C(z)$ of Eq.\,(6) is shown for a sharp
cutoff of $k_{max}=0.4$ GeV. The dotted line represents an
exponential fit to the curve using $m_{sc}=0.961$ GeV. (We recall
that $\pi T$ is equal to 1.27 GeV.)}
\end{figure}

The results of our calculation of $C(z)$ of Eq.\,(6) are given in
Figs. 1 and 2. In Fig. 1 we use $k_{max}=0.1$ GeV and in Fig. 2 we
put $k_{max}=0.4$ GeV. For our calculations, we have $m_{sc}=\pi
T=1.27$ GeV when $T=1.5\,T_c$ and $T_c=0.27$ GeV. Thus, the
$k_{max}$ values considered here are less than $m_{sc}$ and that
feature leads to the exponential behavior seen in Figs. 1 and 2.
If $k_{max}$ is made larger than 0.4 GeV we begin to see
deviations from exponential behavior for $C(z)$. (Since in our
calculations reported in Refs. [1-3], the integrals were regulated
with a Gaussian regulator $\exp[-\vec k\,{}^2/\alpha^2]$ with
$\alpha\simeq4$ GeV, we can see that the $\overrightarrow{k}$
values in those calculations are so large as to preclude obtaining
exponential behavior for our coordinate-space correlator.)

    The goal of this work was to consider a simple quark model for
the calculation of a hadronic current correlation function and to
determine the conditions under which the coordinate-space
correlator is dominated by the screening mass which is given by
the first Matsubara frequency. We have found that the standard
result is obtained if the quark and antiquark momenta in the
vacuum polarization calculation are small compared to that
frequency.



\end{document}